\documentclass[aps,pre,reprint,groupedaddress]{revtex4-2}

\usepackage{amsmath,amssymb,amsfonts,latexsym}
\usepackage{dsfont}
\usepackage{bm} 
\usepackage[mathcal]{euscript}
\usepackage{array}
\usepackage{graphicx}
\usepackage{epsfig}
\usepackage{color}
\usepackage[activate=normal]{pdfcprot}  
\usepackage[colorlinks=true, linkcolor=blue, citecolor=blue]{hyperref}

\newcommand{\gul}{Gulliver UMR CNRS 7083, ESPCI Paris, PSL Research University, 10 rue Vauquelin, 75005 Paris, France}
\newcommand{\isir}{Institut des Systèmes Intelligents et de Robotique, Sorbonne Université, CNRS, ISIR, F-75005 Paris, France.}

\newcommand{\beq}{\begin{equation}}
\newcommand{\eeq}{\end{equation}}
\newcommand{\ben}{\begin{equation*}}
\newcommand{\een}{\end{equation*}}
\newcommand{\bseq}{\begin{subequations}}
\newcommand{\eseq}{\end{subequations}}
\newcommand{\bea}{\begin{eqnarray}}
\newcommand{\eea}{\end{eqnarray}}
\newcommand{\bal}{\begin{align}}
\newcommand{\eal}{\end{align}}

\newcommand{\p}{\partial}

\newcommand{\rb}{\boldsymbol{r}}

\newcommand{\vb}{\boldsymbol{v}}

\newcommand{\Fb}{\boldsymbol{F}}

\newcommand{\Tba}{\boldsymbol{T}_{\rm a}}

\newcommand{\hnb}{\hat{\boldsymbol{n}}}

\newcommand{\heb}{\hat{\boldsymbol{e}}}

\newcommand{\nb}[1]{\begingroup\color[rgb]{0.5,0,0.5}#1\endgroup}

\begin{document}
\graphicspath{{./Figures/}}

\title{Self-Aligning Active Agents with Inertia and Active Torque}

\author{Jeremy Fersula} 
\affiliation{\gul}
\affiliation{\isir}
\author{Nicolas Bredeche}
\affiliation{\isir}
\author{Olivier Dauchot}
\affiliation{\gul}
\date{\today{}}


\begin{abstract}
We extend the study of the inertial effects on the dynamics of active agents to the case where self-alignment is present. In contrast with the most common models of active particles, we find that self-alignment, which couples the rotational dynamics to the translational one, produces unexpected and non-trivial dynamics, already at the deterministic level. Examining first the motion of a free particle, we contrast the role of inertia depending on the sign of the self-aligning torque. When positive, inertia does not alter the steady-state linear motion of an a-chiral self-propelled particle. On the contrary, for a negative self-aligning torque, inertia leads to the destabilization of the linear motion into a spontaneously broken chiral symmetry orbiting dynamics. Adding an active torque, or bias, to the angular dynamics the bifurcation becomes imperfect in favor of the chiral orientation selected by the bias. In the case of a positive self-alignment, the interplay of the active torque and inertia leads to the emergence, out of a saddle-node bifurcation, of truly new solutions, which coexist with the simply biased linear motion. In the context of a free particle, the rotational inertia leaves unchanged the families of steady-state solutions but can modify their stability properties. The situation is radically different when considering the case of a collision with a wall, where a very singular oscillating dynamics takes place which can only be captured if both translational and rotational inertia are present.
\end{abstract}


\maketitle


\section{Introduction}
\label{sec:intro}
Self-propelled agents, the "big atoms" of active matter, consume energy to produce directed motion. In many cases, such as bacteria~\cite{Wu-2000,Dombrowski-2004,Zhang-2010,Peruani-2012}, cells~\cite{Szabo-2006a,Smeets-2016,Peyret-2019}, man-made Janus~\cite{Walther-2008,Palacci-2010,Palacci-2013,Zheng-2013,Buttinoni-2013,Ginot-2015} or rolling colloids~\cite{Bricard-2013,Geyer-2018}, the agent size together with the viscosity of the surrounding medium ensure that the dynamics take place at sufficiently low Reynolds number and inertia can be neglected~\cite{Bechinger-2016}. 

There are however other cases, where this simplification does not hold, as for instance with the flight of birds~\cite{Cavagna-2014,Attanasi-2014}, the motion of vibrated polar grains~\cite{Kudrolli-2008,Deseigne-2010a,Weber-2013a,Kumar-2014}, or that of centimetric robots~\cite{Giomi-2012,Ferrante-2012,Rubenstein-2014,ben2023morphological}. 
An important effort has been made to upgrade the model of active Brownian particles (ABP) by including inertia in both translational and orientational motion~\cite{Scholz-2018,Lowen-2020}. Analytical results were obtained for the orientational and translational correlation functions of the single particle dynamics with good agreement with experimental results from vibrated granular systems.
More exact analytical predictions for higher-order statistics were also obtained in devising an inertial Active Ornstein-Uhlenbeck particle (AOUP), which further simplifies the ABP dynamics by enforcing Gaussian fluctuations~\cite{Sprenger-2023}. More specifically, it was shown that rotational inertia is fundamentally relevant to reproduce the temporal delay between the active force and particle velocity observed for a single active granular particle.

Another important ingredient to describe polar agents, especially those which take their momentum from a substrate, is self-alignment. Self-alignment was introduced as early as 1996' in the pioneering work of~\cite{Shimoyama-1996a}, stemming from the very basic observation that the heading and the velocity of a polar body do not need to be parallel.
When they are not, the distribution of propulsive and dissipative forces is generically not symmetric concerning the body axis and therefore exerts a torque on the agent body; this is self-alignment. 
It was reintroduced independently in~\cite{Szabo-2006} to describe the collective migration of tissue cells, in~\cite{Henkes-2011} to study active jamming, in~\cite{Ferrante-2012} to describe an assembly of wheeled robots and in~\cite{Deseigne-2010, Weber-2013, NguyenThuLam-2015a}, where it was shown to be a key ingredient for the emergence of collective motion in a system of self-propelled polar disks. 
More recently, it has started to attract more attention in the context of dense and solid active matter. It was introduced in vertex models~\cite{Malinverno-2017, Barton-2017, Giavazzi-2018, Petrolli-2019}, in phase field models~\cite{Peyret-2019}, and in a model experimental system of active elastic networks where its central role was elucidated~\cite{Baconnier-2022,Baconnier-2023}. Finally, self-alignment was recently used as a morphological asset in the context of swarm robotics~\cite{ben2023morphological}.

\begin{figure*}
\center
\includegraphics[width = 0.95\textwidth]{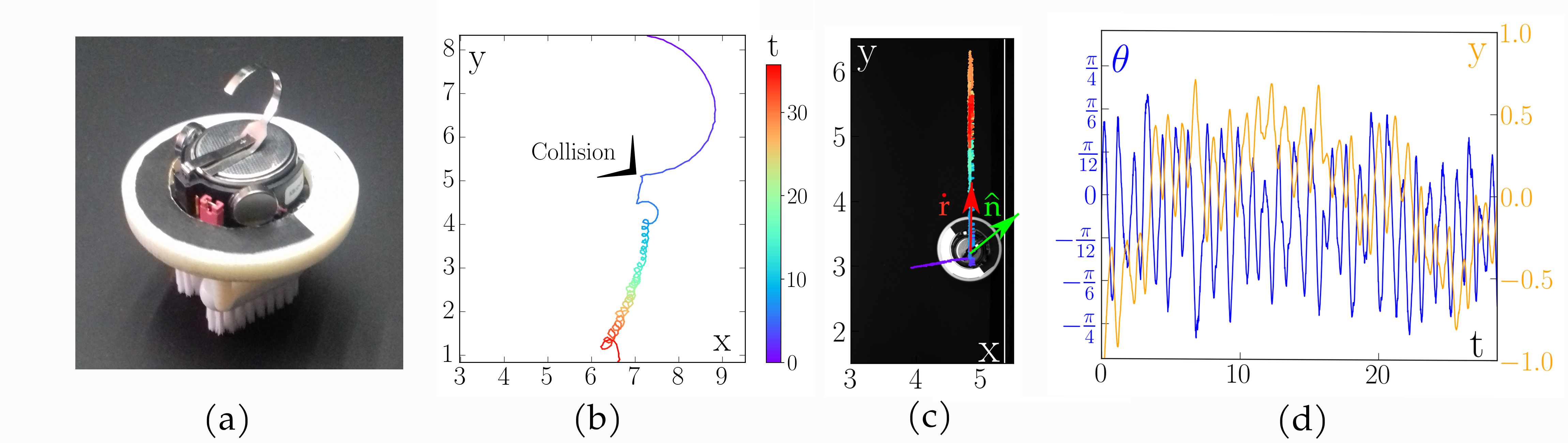}
\caption{{\bf A few instances of unexpected dynamics observed with an inertial self-aligning agent:} (a) 
A Vibebot, combining a standard Kilobot platform~\cite{Rubenstein-2014} with a custom-designed exoskeleton, featuring toothbrush legs in place of the original three metallic rod legs; (b) such a robot, moving at 10 cm/s is kicked by an external perturbation, while performing a rather large gyration radius circular trajectory, and switches to a rapidly spinning dynamics (see also Supp. Mat. Movie 1); when such augmented Kilobot collisions with a wall with small enough velocity (c), it oscillates near its impact point, reorienting periodically (d) (see also Supp. Mat. Movie 2); eventually, when the incoming velocity is large and the oscillation amplitude is large enough, the robot escapes the wall (see also Supp. Mat. Movie 3). }
\label{fig:exp}
\end{figure*}

In the absence of self-alignment, the angular dynamics decouples from the translational motion. As a result, the steady-state deterministic dynamics are trivial. For an a-chiral, that is a left-right symmetric agent, it is linear motion; for a chiral or left-right biased agent, it is circular motion. The research activity has therefore concentrated on the description of the statistics of the fluctuating motion in the presence of translational and/or rotational noise~\cite{Scholz-2018,Lowen-2020,Sprenger-2023}. 
In a recent attempt to design new low-cost robots, we noted the emergence of new dynamics, reported in Fig.~\ref{fig:exp}. Following a collision, or a manual perturbation, a single freely moving robot can abruptly switch from a circular-like motion to a spinning one (Fig.~\ref{fig:exp}-b), revealing an unexpected coexistence of two very different dynamics for the same value of the control parameters. Collisions with a linear wall reveal even more curious behavior, with the robot facing the wall while performing a peculiar angular and translational oscillating motion, around its point of impact (Fig.~\ref{fig:exp}-c,d). The amplitude of the oscillations increases with the incoming speed, eventually reorienting the robot away from the wall. Such qualitatively new and intriguing dynamics must emerge from the coupling of translational and rotational motions through self-alignment.

The main goal of this paper is to decipher the respective role of bias and inertia in setting up these dynamics for self-aligning polar agents. To do so we analyze the effect of translational and rotational inertia on the deterministic dynamics of a self-aligning polar particle, with or without an additional angular bias.
Note that the robot used in the above discussion being a prototype, we don't aim here at extracting the precise parameters of the experimental system to quantitatively reproduce the dynamics in silico.  
The paper is organized into two main parts.  We first discuss the case of a free particle, before considering the interaction with a linear wall. 
In the first part, we show that the role of translational inertia depends on the sign of the self-aligning torque. When positive, inertia does not modify the steady-state linear motion of an a-chiral self-propelled particle. However, its interplay with an additional angular bias, leads to the emergence, out of a saddle-node bifurcation, of truly new solutions, which coexist with the simply biased linear motion.  On the contrary, for a negative self-aligning torque, inertia leads to the destabilization of the linear motion into a spontaneously broken chiral symmetry orbiting dynamics. An additional bias simply turns the bifurcation into an imperfect one in favor of the chiral orientation selected by the bias. As long as a free particle is considered, the rotational inertia leaves unchanged the families of steady-state solutions, except for their linear stability. 
In the second part, we show that the situation is radically different when considering the case of a collision with a wall, where the very singular oscillating dynamics described above can only be captured if both translational and rotational inertia are present.

Our starting point is given by Newton's equations describing the deterministic motion of a self-aligning polar agent heading along $\hnb$ in two dimensions: %
\bseq \label{eq:Inertia}
\begin{align} 
	m \ddot\rb &= F_a \hnb - \gamma \dot\rb + \tilde\Fb_{\rm ext}(\rb), \label{eq:Inertia_trans}\\
	\tilde J   \ddot{\hnb} 	&=  \tilde\Tba \times \hnb + \zeta (\hnb \times \dot\rb) \times \hnb  - \gamma_r \dot{\hnb}.  \label{eq:Inertia_rot}
\end{align}
\eseq
\nb{}

where $m$\nb{, $r$} and $\tilde J$ respectively are the mass\nb{, position} and inertial momentum of the agent, and $\gamma$, $\gamma_r$ respectively encode the translational and rotational damping, which are assumed to be scalar. The first equation describes the inertial translational motion of an active agent self-propelled by an active force $F_a \hnb$ and subjected to an external force $\tilde\Fb_{\rm ext}(\rb)$. The second equation expresses the inertial reorientation of the active force, subject to an active torque $\tilde\Tba$ and self-alignment. The active torque is perpendicular to the particle motion and models a possible chirality of the active particle, also called bias if undesired. The self-aligning term expresses the coupling between the velocity of the particle and the orientation of the active force when they are not colinear. When $\zeta$, the amplitude of the self-aligning coupling, is positive, respectively negative, the coupling tends to align, resp. anti-align, $\hnb$ with $\dot\rb$.

These equations were shown to faithfully describe the motion of a self-propelled polar agent in a harmonic potential, as experimentally observed with the simple Hexbug\copyright~ robot device running in a parabola dish~\cite{Dauchot-2019}. They also capture the onset of collective motion in a system of self-aligning hard disks~\cite{Lam-2015,Lam-2015a}, as observed experimentally and numerically in a system of vibrated polar grains~\cite{Deseigne-2010a,Deseigne-2012a,Weber-2013a}. Note that in some other context~\cite{Shimoyama-1996a,Szabo-2006a,Henkes-2011a} the self-aligning torque can be normalized by the norm of $\dot\rb$.

In the following, we shall respectively use $m$, the body length $d$ of the agent, and $d/v_0$, with $v_0 = F_a/\gamma$ the free flight velocity, as the mass, length and time units. The dimensionless equations then read :
\bseq \label{eq:Inertia_nodim}
\begin{align} 
	\tau_v \ddot\rb &=  \hnb -  \dot\rb + \Fb_{\rm ext}(\rb), \label{eq:Inertia_trans_nodim}\\
	J   \ddot{\hnb} 	&=  \Tba \times \hnb + \epsilon (\hnb \times \dot\rb) \times \hnb  - \tau_n \dot{\hnb}.  \label{eq:Inertia_rot_nodim}
\end{align}
\eseq
with $\Fb_{\rm ext} = \frac{\tilde \Fb_{\rm ext}}{\gamma v_0}$ and $\Tba= \frac{\tilde\Tba}{|\zeta| v_0}$, $J = \frac{\tilde J v_0}{|\zeta| d^2}$, $\tau_v = \frac{m v_0}{\gamma d}$, $\tau_n = \frac{\gamma_r}{|\zeta| d}$ and $\epsilon = sign(\zeta)$.

\section{Free particle dynamics}
\label{sec:free}

In the absence of external force, the isotropy of space imposes that only the difference of orientation between $\hnb$ and $\vb = \dot\rb$ matters. Introducing the orientations $\phi$ of $\vb = v \begin{pmatrix} \cos \phi \\ \sin \phi \end{pmatrix} $ and $\theta$ of $\hnb = \begin{pmatrix} \cos \theta \\ \sin \theta \end{pmatrix}$, together with their difference $\alpha = \theta - \phi$, one obtains the equations for the free particles dynamics :
\bseq \label{eq:free_nodim}
\begin{align} 
    \tau_v \dot{\phi} &= \frac{1}{v} \sin\alpha \label{eq:free_nodim-a}\\
    \tau_v \dot{v} &= \cos\alpha - v \label{eq:free_nodim-b}\\ 
    \dot{\alpha} &= \omega -  \frac{1}{\tau_v v} \sin\alpha \label{eq:free_nodim-c}\\
    J \dot{\omega} &= -\epsilon v \sin\alpha - \tau_n  \omega + T_a \label{eq:free_nodim-d}
\end{align} 
\eseq
where $\omega = \dot\theta$. The last three equations form a closed system for the variable $(v, \alpha, \omega)$, the solution of which sets the dynamics of $\phi$ through the first equation.

The steady-state dynamics are obtained by solving for the fixed points of equations~(\ref{eq:free_nodim}) and performing their linear stability analysis.
Equations~(\ref{eq:free_nodim-b}) and~(\ref{eq:free_nodim-c}) readily lead to $v^* = \cos\alpha^*$ and $\omega^* = \frac{1}{\tau_v}\tan\alpha^*$. Substituting in eq.~(\ref{eq:free_nodim-d}) and denoting $t=\tan\alpha^*$, one finds the third order polynomial in $t$, the roots of which sets the fixed points:
\begin{equation}
\label{eq:poly_fp}
\frac{\tau_n}{\tau_v} t^3 -T_a t^2 + t(\epsilon +\frac{\tau_n}{\tau_v}) - T_a = 0
\end{equation}

\subsection{Unbiased inertial dynamics}
\begin{figure}[t!]
    \centering
    \includegraphics[width=0.9\linewidth]{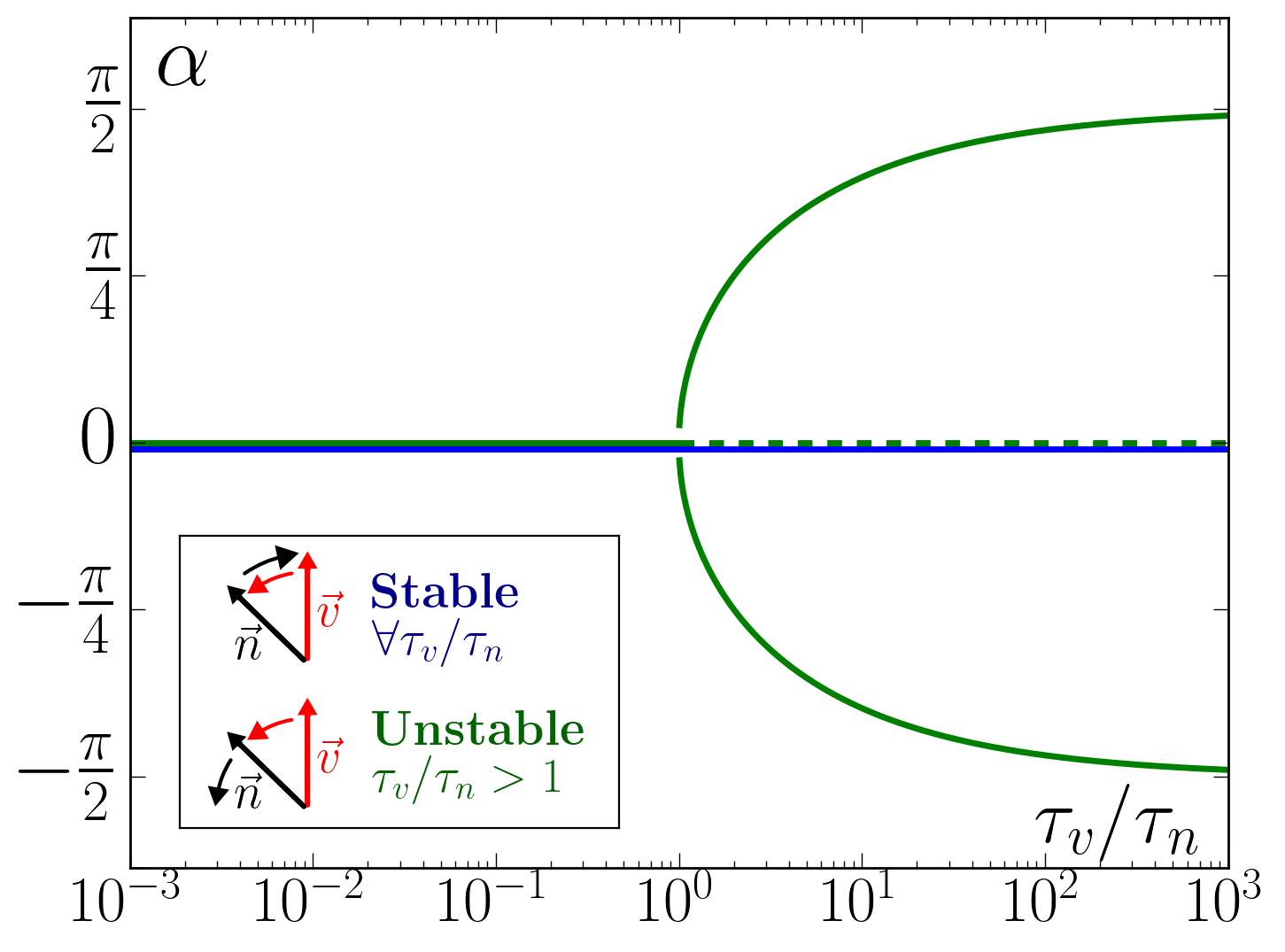}\\
    (a)\\
    \includegraphics[width=0.9\linewidth]{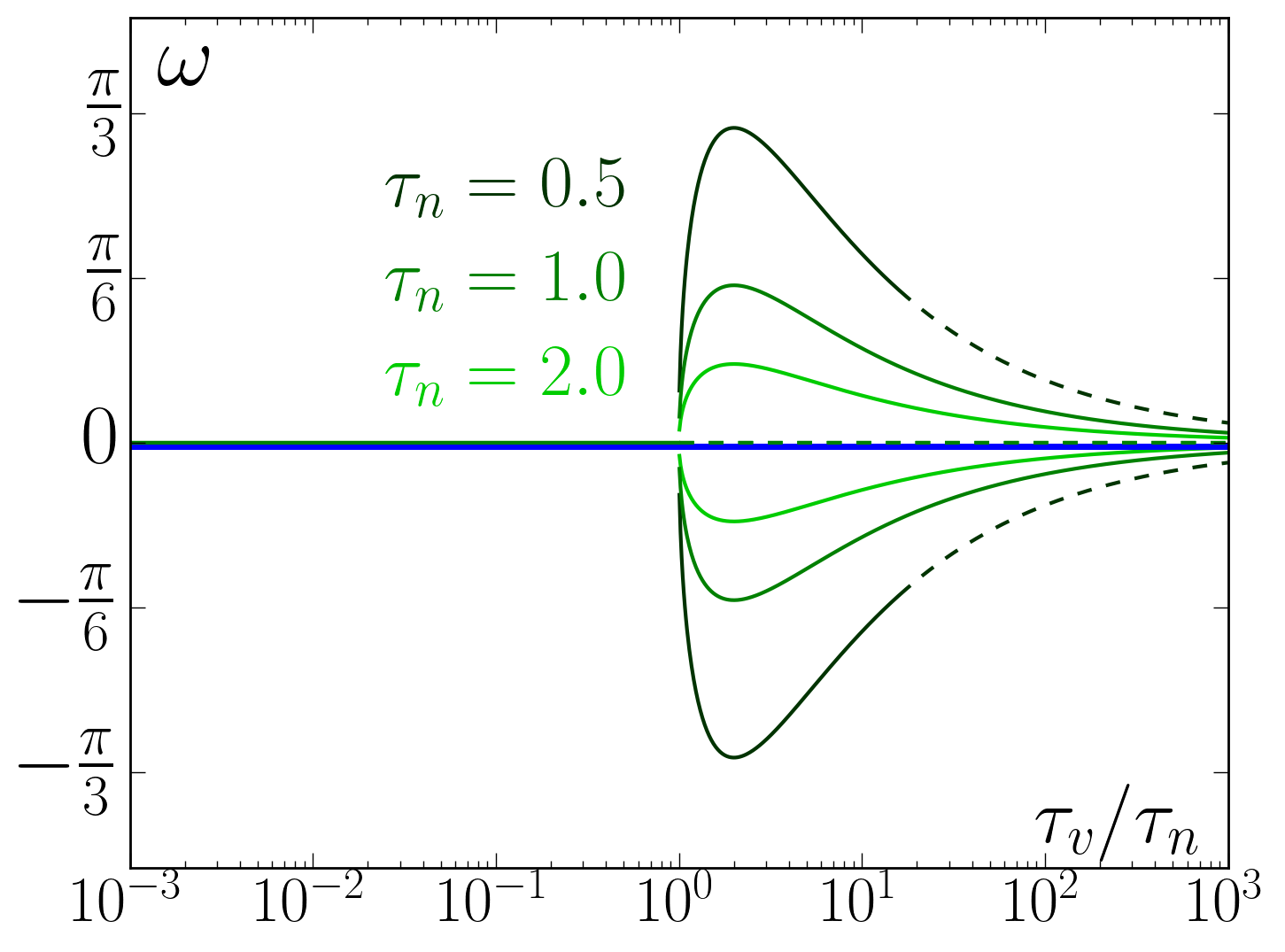}\\
    (b)
    \caption{{\bf Bifurcation diagram for the steady state dynamics of a free self aligning particle with inertia and no bias:} (a) angle between the orientation and the velocity vector; Inset: while the aligning dynamics is stabilizing the dynamics, the anti-aligning one is destabilizing; (b) orbiting frequency. In blue, resp. green: aligning, resp. anti-aligning case. Continuous, resp. dotted lines denote linearly stable, resp. unstable, solutions.}
    \label{fig:bif_nobias}
\end{figure}

It is instructive to start with the case of an a-chiral, or unbiased, particle, $T_a=0$,  for which obtaining the steady state solutions and their stability is straightforward. The results are summarized in Fig.~\ref{fig:bif_nobias}.
One immediately identifies the trivial fixed point $t=0$, leading to $v^* = 1, \alpha^* = 0, \omega^* = 0, \dot\phi^* = 0$, which corresponds to the particle performing straight motion at nominal velocity, with $\vb$ and $\hnb$ being aligned. 
In the aligning case, $\epsilon = +1$ (blue lines in  Fig.~\ref{fig:bif_nobias}), this is the only fixed point and it is always linearly stable.
In the anti-aligning case, $\epsilon = -1$ (green lines in  Fig.~\ref{fig:bif_nobias}), two mirror fixed points, given by  $\tan\alpha^*=\pm \sqrt {- (1 + \epsilon  \tau_v/\tau_n))}$ emerge from a a pitchfork bifurcation, when the linear motion turns unstable for $\tau_v/\tau_n >1$. These fixed points describe clockwise and anti-clockwise circular trajectories, with an orbiting frequency $\omega^*= \dot\phi^* =  \frac{1}{\sqrt{\tau_n \tau_v}}\sin\alpha^*$, and a radius $R^*=v^*/\omega^*= \frac{1}{\tau_v \sin\alpha^*}$. Note the non monotonic dependence of the rotation frequency with $\tau_v\tau_n$: it arises from the combination of the fast increase of $\alpha^*$ at the onset of the instability and the $1/\sqrt{\tau_v}$ prefactor expressing the slowing down of the dynamics by inertia

The physical picture is as follows. In both cases the velocity $\vb$ tends to align with the orientation $\hnb$ imposed by the active force. In the aligning case, $\hnb$ rotates toward $\vb$, so that independently of the timescales of the dynamics, the two vectors dynamics stabilize the steady state where they are aligned. Conversely, in the anti-aligning case,  $\hnb$ rotates \emph{away} from $\vb$. Whenever $\tau_v/\tau_n$ is too large, $\vb$ cannot catch up with $\hnb$, and any small disturbance of the linear motion destabilizes it in an endless orbiting motion. 
The angular inertia does not alter the family of solutions, but as we shall discuss below in the general case with, it can modify the stability properties of these solutions. 
We conclude this section by recalling that, to our knowledge, there is so far only one experimental realization of an anti-aligning polar particle, a kilobot augmented with a specific 3d exoskeleton described in~\cite{BenZion-2023}. The orbiting solution pinpointed here was not reported in that work, the reason being that the inertia of such robots is small enough to avoid the destabilization of the straight motion. 

\subsection{Inertial dynamics of chiral particles}
\begin{figure}[t!]
\centering
\includegraphics[width=0.9\linewidth]{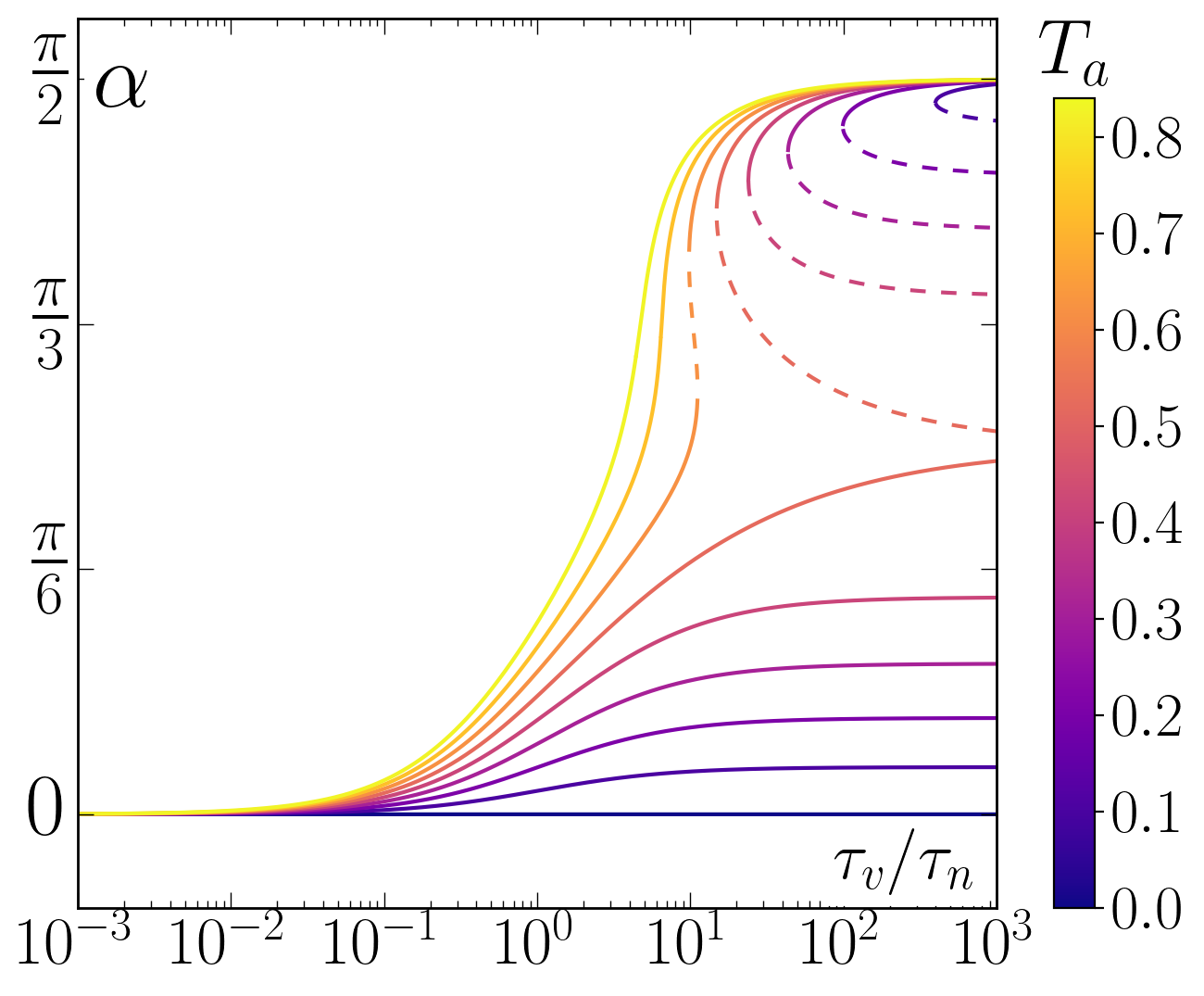}\\
    (a)\\
\includegraphics[width=0.9\linewidth]{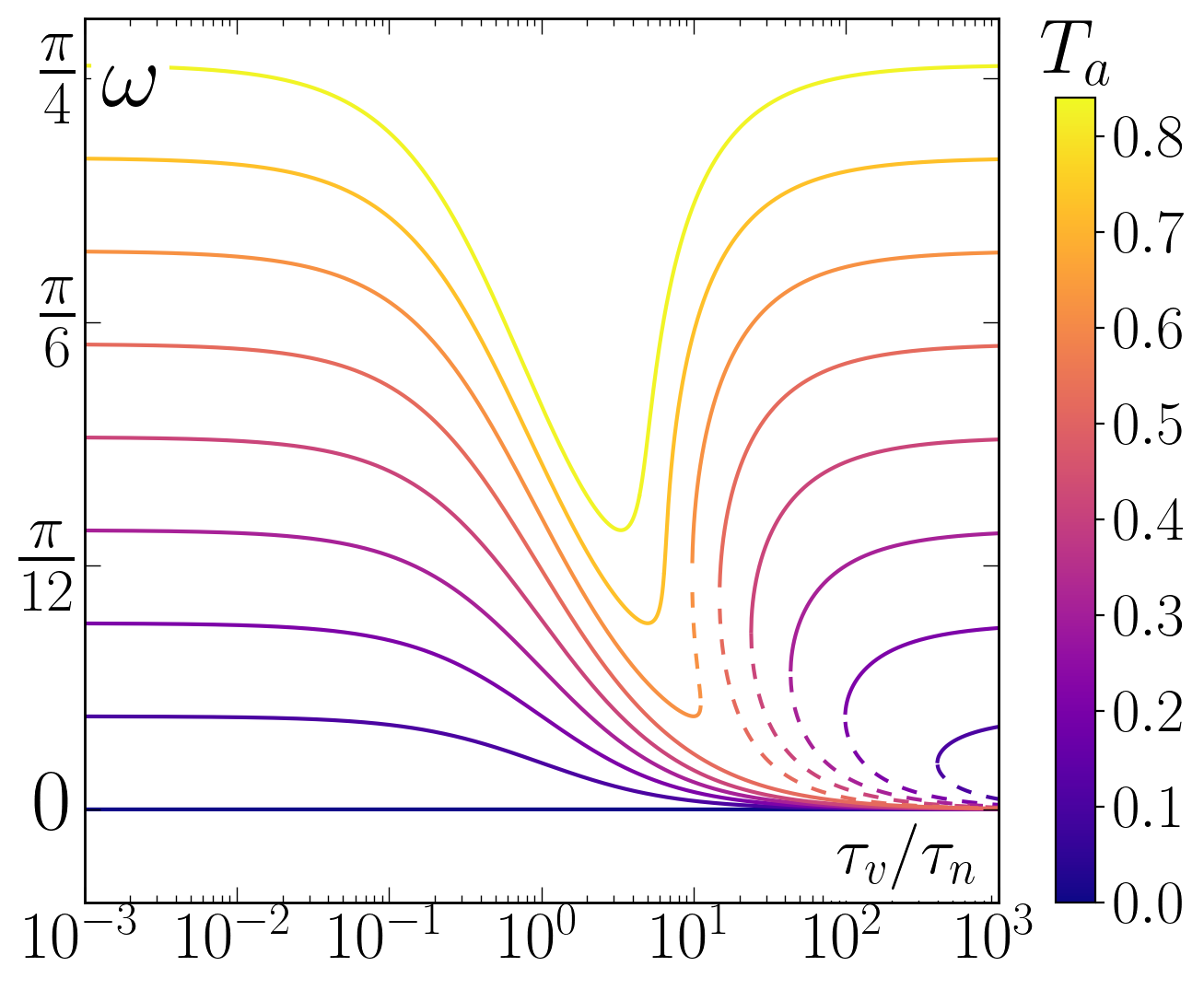}\\
    (b)
\caption{{\bf Free inertial dynamics of aligning chiral particles:} (a) Misalignment, $\alpha^*$, between the velocity and the orientation of the self propelling force and (b) angular frequency, $\omega^*$, of the resulting circular motions as a function of $\tau_v/\tau_n$ for increasing values of the bias, as indicated by the color code. Continuous, resp. dotted lines denote linearly stable, resp. unstable, solutions.  }
\label{fig:AlignerBifurcation}
\end{figure}

\begin{figure}[t!]
\centering
\includegraphics[width=0.9\linewidth]{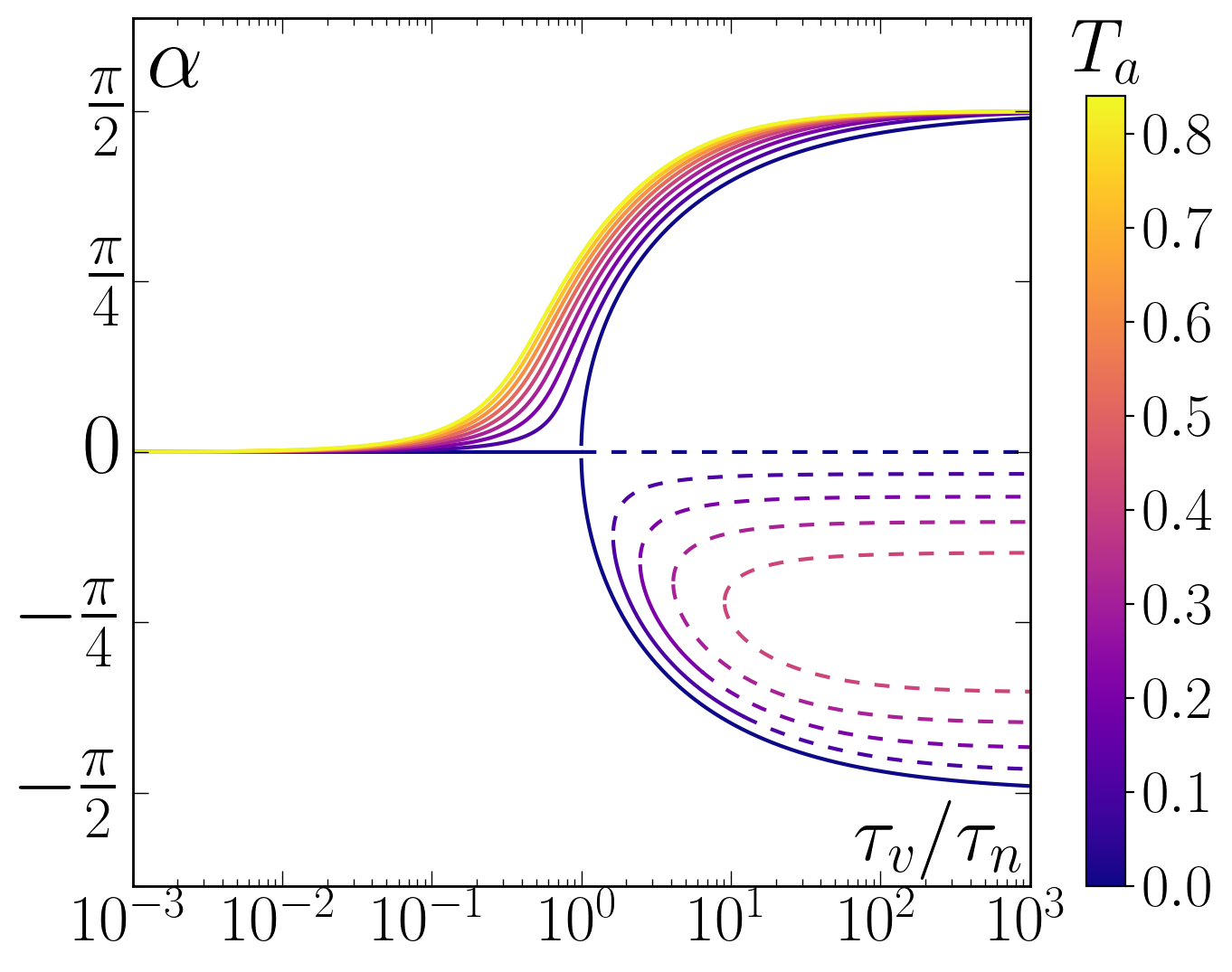}\\
    (a)\\
\includegraphics[width=0.9\linewidth]{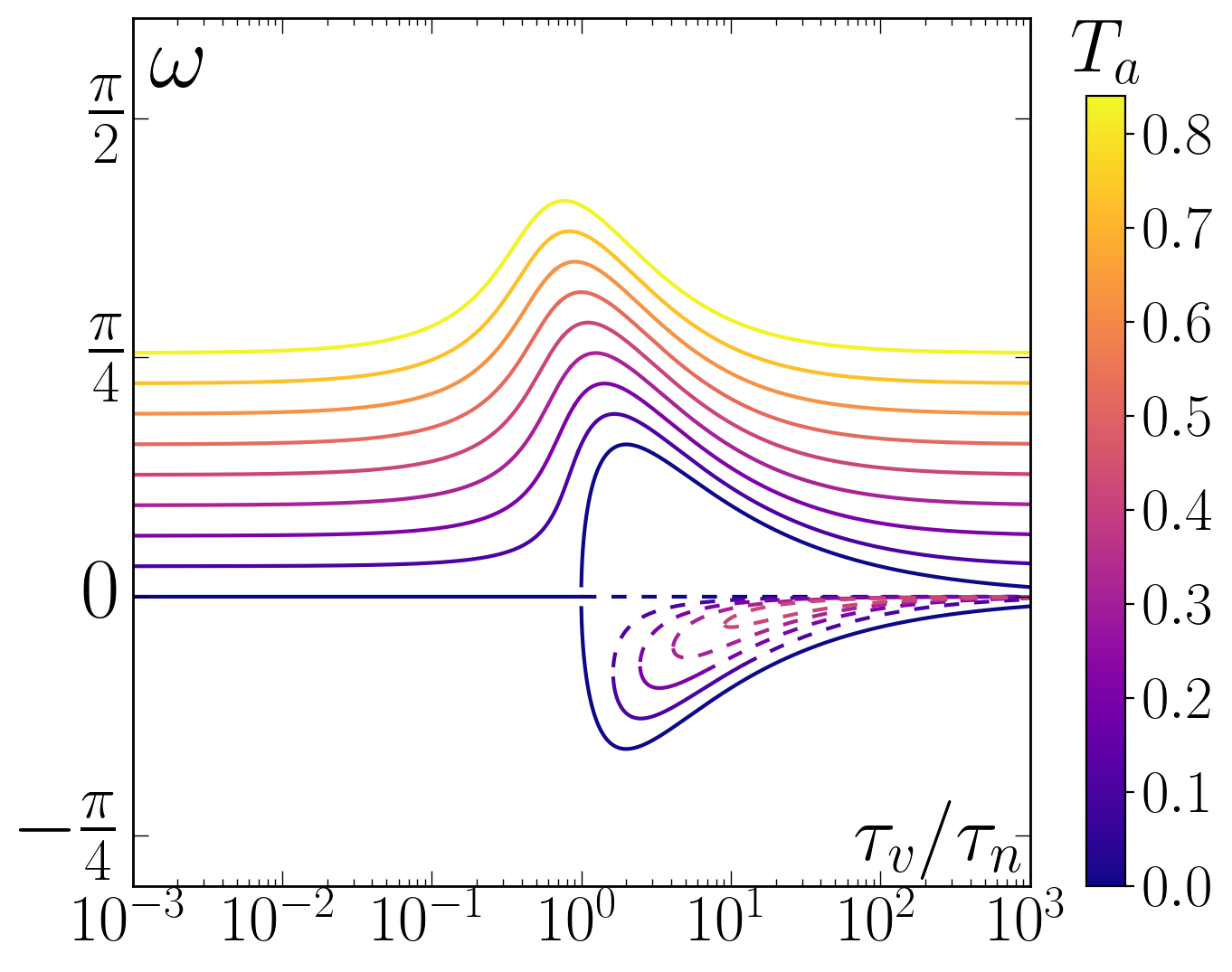}\\
    (b)\\
\caption{{\bf Free inertial dynamics of an anti-aligning chiral particle:} (a) Misalignment, $\alpha^*$, between the velocity and the orientation of the self propelling force and (b) angular frequency, $\omega^*$, of the resulting circular motions as a function of $\tau_v/\tau_n$ for increasing values of the bias, as indicated by the color code. Continuous, resp. dotted lines denote linearly stable, resp. unstable, solutions. }
\label{fig:FronterBifurcation}
\end{figure}

In presence of an active torque, the analytical solutions provided by solving for the roots of eq.~(\ref{eq:poly_fp}) have a cumbersome dependence on the parameters and are provided in the appendix. However one can make a few simple statements by considering limiting cases. 
In the limit of vanishing translational inertia,  $\frac{\tau_v}{\tau_n} \rightarrow 0$, the only solution is $\alpha^* \rightarrow 0$, $v^*\rightarrow 1$ and $\omega^*\rightarrow T_a / \tau_n$. It corresponds to circular trajectories with a radius $R^* \rightarrow \tau_n / T_a$, which diverges in the a-chiral limit: they simply form the generalization of the straight trajectories bended by the bias.
Conversely, in the limit of large translational inertia, $\frac{\tau_v}{\tau_n} \rightarrow \infty$, the situation becomes qualitatively different. For $\left|T_a\right| < 1/2$, three solutions exists, one of which being given by  $\alpha^* \rightarrow \pm \frac{\pi}{2}$, 
$v^*\rightarrow 0$ and $\omega^* \rightarrow T_a / \tau_n$. For $\left|T_a\right| > 1/2$, only this solution subsists. It correspond to a purely spinning dynamics, where the particle rotates on itself.
The connection between the two limits is summarized on Figs.~\ref{fig:AlignerBifurcation} and Figs.~\ref{fig:FronterBifurcation} for the aligning and anti-aligning case respectively.


In the aligning case, $\epsilon = +1$, one finds an unexpectedly rich bifurcation diagram.  As expected the linear motion obtained when $T_a=0$ is replaced by the slowly rotating circular motion, with $\omega^* = \frac{\tan\alpha^*}{\tau_v}$, the radius of which $R^*=\frac{1}{\tau_v \sin\alpha^*}$ decreases from infinity when the bias grows from zero. This solution is always linearly stable. More surprising are the two solutions, which emerge from a saddle node bifurcation above a bias-dependent critical value of $\tau_v/\tau_n$, when $T_a<1/2$. The fastest one, with a small radius of gyration, converges to the spinning solution described above in the limit of large inertia $\tau_v/\tau_n \rightarrow \infty$ and is stable. The slower one is always unstable. When $T_a>1/2$, the lower branch disappears and the faster branch merges with the slowly rotating circular motion inherited from the straight motion, in the absence of bias.

Comparatively, the anti-aligning case,  $\epsilon = -1$, is a straightforward generalization of the bifurcation diagram obtained in the absence of bias.  For $T_a <1/2$, the active torque simply renders the pitchfork bifurcation imperfect, in a way analogous to the effect of an external field on a para-ferromagnetic transition. The bifurcated branch compatible with the bias merges continuously with the slowly rotating circular motion inherited from the straight motion, in the absence of bias. The other one connects in a saddle-node bifurcation to the one inherited from the linearly unstable straight trajectory. For $T_a >1/2$, only the linearly stable strongly biased solution persists.

\begin{figure}
\centering
\includegraphics[width=0.9\linewidth]{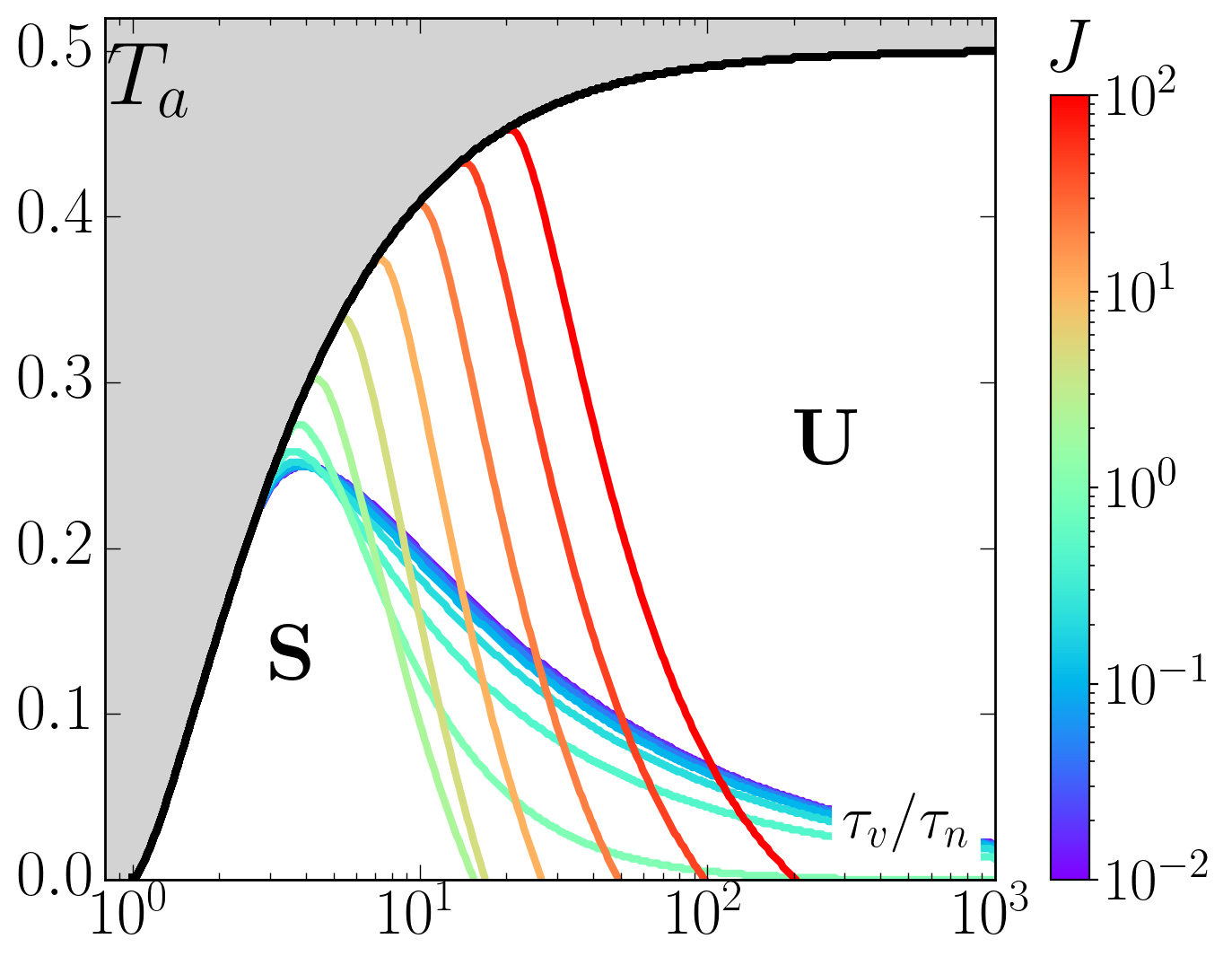}
\caption{{\bf Domain of existence and stability of the lowest branch of the bifurcated solutions, in the case $\epsilon = -1$:} The grey area denotes the region in the ($\tau_v/\tau_n, T_a$) plane, where the solution does not exist. In the existence domain, the solution is stable to the left of a line, the position of which is set by the angular inertia, as indicated by the color code  ($\tau_n = 1$).}
\label{fig:taunJStability}
\end{figure}

We close this section by discussing the non-trivial dependence on the angular inertia of the linear stability of the bifurcated solutions in the anti-aligning case. The solution inherited from the linearly unstable straight trajectory remains unstable as it should. The linear stability of the solution inherited from the bifurcated solution opposing the bias on the contrary depends on the angular inertia $J$ and $\tau_n$.  Fig.~\ref {fig:taunJStability} displays the existence and stability domains of these solutions, in the ($\tau_v/\tau_n, T_a$) plane for $\tau_n = 1$ and different values of $J$. As stated above, the bias sets the domain of existence of the solutions: the larger $T_a$, the larger the value of $\tau_v/\tau_n$ above which these solutions exist, before eventually disappearing for $T_a>1/2$. A finite angular inertia imposes a bias-dependent maximal value to $\tau_v/\tau_n$, above which the solution turns linearly unstable. For small $J$, the so-obtained domain of stability shrinks when $J$ increases. For $J$ larger than a threshold of the order of $\tau_n$, this tendency reverses: the linear stability enlarges with growing $J$, eventually recovering stable solutions in all their domain of existence in the limit $J \rightarrow +\infty$. A similar dependence is observed for an increase of $1/\tau_n$ at a fixed value of $J$, highlighting the similar role played by the two quantities. The detailed linear analysis is provided in appendix.

In conclusion, the above results demonstrate the key role of self-alignment, when it combines with translational inertia, in producing qualitatively new steady dynamics even in the simplest case of freely moving particles. For the aligning case, the straight-line trajectory remains the only solution in the absence of bias. Any small amount of bias gives rise to a new branch of solution disconnected from the previous one, that corresponds to a rapidly spinning motion. Such a solution coexists with the quasi-linear motion. For the anti-aligning case, two branches of chiral orbiting solutions emerge at large inertia even in the absence of bias. 

In the absence of external forces, the rotational inertia only affects the stability of these new steady dynamics.
We shall now see that it becomes a central ingredient when describing the interaction of such a particle with a linear hard wall. 

\section{Collision with a wall}
\label{sec:wall}

Performing experiments with self-propelled agents, one usually confines them in one way or another, typically using lateral walls. Fig.~\ref{fig:exp}(c-e) reports the motion of an inertial aligning agent, when it enters into contact with a wall. For low enough incoming velocity, one observes that, after a short transient, the self-aligning agent performs an endless translational and orientational oscillation along the wall. The amplitude of these oscillations increases with the incoming speed. When the amplitude of the angular oscillation is so large that it reorients the agent away from the wall, the latter escapes the wall.

This intriguing dynamics is well captured by equations (\ref{eq:Inertia_nodim}) where the external force now describes the interaction with the wall. As we shall see the bias is not a necessary ingredient and we omit it from now on, for the purpose of simplicity.  Let us denote $\heb_x$ the direction towards the wall and $\heb_y$, the direction parallel to the wall (see Fig.~\ref{fig:exp}-c) with the wall sitting in the position $x=0$. In the light of the rolling motion observed experimentally, the most general description of this interaction is that of 
\begin{itemize}
\item a repulsive force in the direction normal to the wall: $ F^{\perp}_w = - \frac{\p V_w(\rb)}{\p x} = - \frac{\epsilon_w}{d} N_w$, with $- N_w$ being the dimensionless force normal to the wall;  
\item a tangential frictional force opposing the sliding velocity $u$: $ F^{\parallel}_w = -\mu u$, with $u=v_y + \frac{d}{2}\dot\theta$, where the second term accounts for the rotation of the contact point with the wall; 
\item a torque resulting from the friction at contact: $\Gamma_w = -\mu \frac{d}{2} u$.
\end{itemize}
The precise choice of the repulsive potential is not crucial as long as it is stiff enough. In the following, $V_w$ is a Weeks-Chandler-Andersen (WCA) potential, leading to a normal force $$N_w = 16 n \left(\left(\frac{\sigma}{|x|}\right)^{2n+1} - \frac{1}{2} \left(\frac{\sigma}{|x|}\right)^{n+1} \right),$$ with $\sigma = d/2^{1+1/n}$ and $n=6$. Finally, the interaction with the wall is truncated and set to zero when the distance to the wall $|x|> 2^{1/n} \frac{d}{2}$.
Altogether the dimensionless equations describing the dynamics of a self-aligning agent in contact with the wall read:
\bseq \label{eq:Wall_nodim}
\begin{align} 
	\tau_v \ddot{x} &=  \cos\theta - \dot x  - \kappa N_w(x), \label{eq:Wall_x_nodim}\\
	\tau_v \ddot{y} &=  \sin\theta -  \dot y  - \nu \left(\dot y + \dot\theta/2 \right), \label{eq:Wall_y_nodim}\\
	J   \ddot{\theta} 	&=  \epsilon \left(\cos\theta\,\dot{y} - \sin\theta\,\dot{x} \right) - \tau_n \dot{\theta} - \tau_r \left(\dot y + \dot\theta/2 \right),  \label{eq:Wall_rot_nodim}
\end{align}
\eseq
where $\tau_v$, $J$, $\tau_n$, $\epsilon$ are as defined in the previous section and the additional dimensionless parameters are $\kappa = \frac{\epsilon_w}{\gamma v_0 d}$, $\nu = \frac{\mu}{\gamma}$ and $\tau_r=\frac{\mu d}{2 \zeta}$.

As a further simplification, we shall assume that the relevant part of the dynamics takes place along the wall, while the dynamics perpendicular to the wall consist of a rapid equilibration of the propelling force and the repulsive one. This is possible because the wall is not infinitely rigid and the position "within" the wall can accommodate for the variation of the propelling force in the direction normal to the wall according to the balance $\cos\theta = \kappa N_w(x)$. We shall verify below that this assumption is valid, by performing simulations of the above equations, once a better understanding of the mechanisms at play, will allow us to select the proper range of values for the numerous control parameters.
\begin{figure}[t!]
\centering
\includegraphics[width=0.99\linewidth]{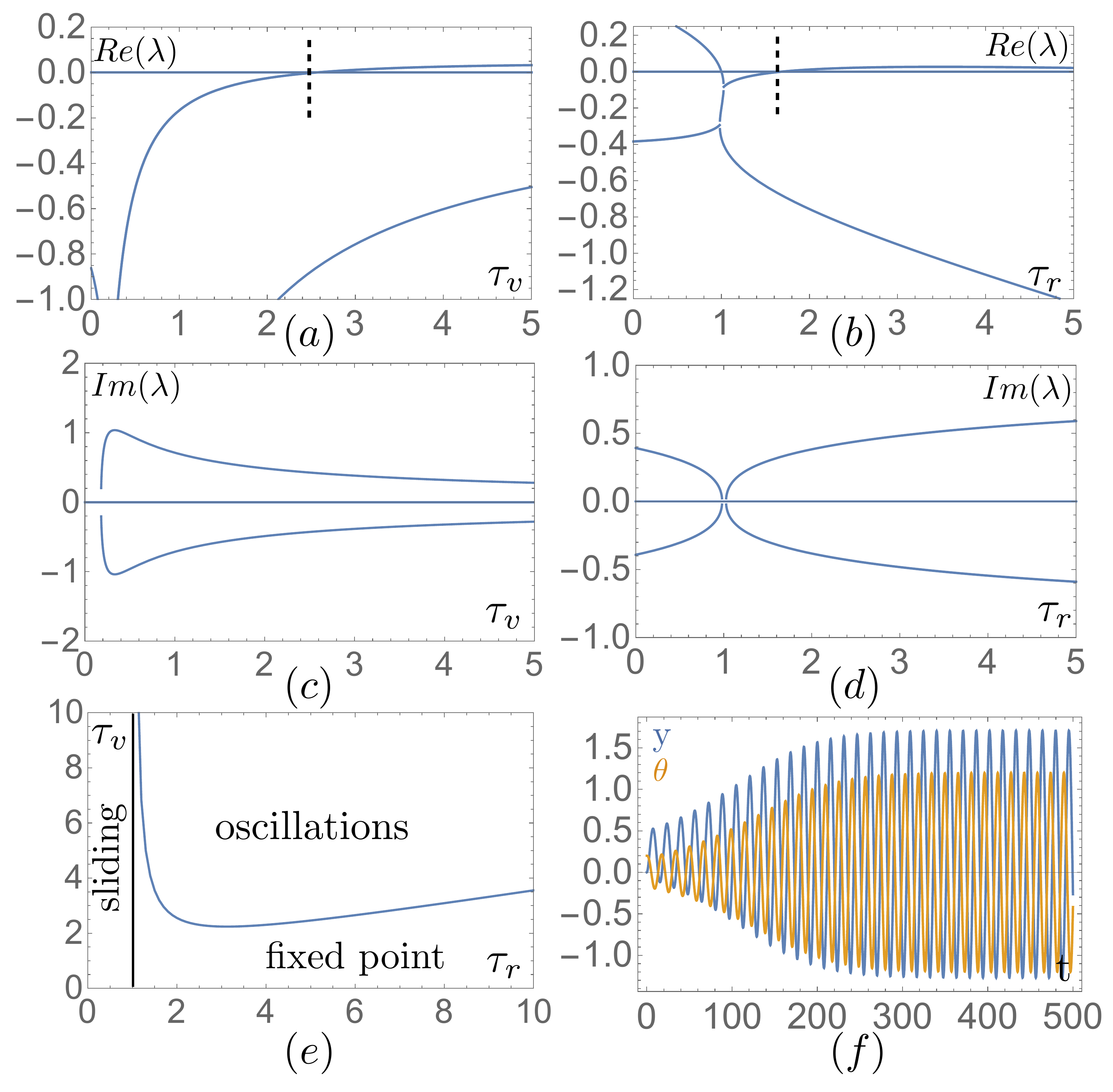}
\caption{{\bf Linear stability of the fixed point against the wall in the aligning case, $\epsilon = +1$:} (a-c) Real and imaginary part of some, including the largest, eigenvalues as a function of $\tau_v$ for $\tau_r=2$. For large enough inertia, the fixed point facing the wall turns unstable via a Hopf bifurcation (vertical dotted line); (b-d) Real and imaginary part of the eigenvalues as a function of $\tau_r$ for $\tau_v=3$ : for $\tau_r<1$, there is always a positive real part and the fixed point is unstable; for $\tau_r>1$ there is a small range of $\tau_r$ values for which the fixed point is stable, before it turns unstable via a Hopf bifurcation (vertical dotted line).  (e) The critical value of $\tau_v$ above which oscillations take place depends in a non-monotonous way on $\tau_r$. (f) The oscillating dynamics for $\tau_v=3$ and $\tau_r=2$. In all panels $J = \tau_v,  \tau_n = 0.1, \nu = 0.1$.}
\label{fig:wall_align}
\end{figure}
In the aligning case, $\epsilon = +1$, the dynamics described by eqs.~(\ref{eq:Wall_nodim}), reduced to the dynamics along $y$ and $\theta$, have two infinite sets of fixed points $(y^*=y_0, \theta^* = 0\, {\rm or}\, \pi, \dot{y}^* = 0, \dot{\theta}^* = 0)$ corresponding to the active agent pointing statically into or outward the wall at any position $y_0$, the case where the agent points outward being irrelevant here. In the absence of inertia, these fixed points are marginally stable, reflecting the translational invariance along $y$. For any small amount of inertia, any of these fixed points becomes linearly unstable for $\tau_r<1$ (Fig.~\ref{fig:wall_align}-b,d) and the dynamics obey another steady solution that is given by $(y_s = v_0 t, \theta_s = \arccos(\tau_r), \dot{y}_s = v_0 = \sin\theta_s/(1+\lambda), \dot\theta_s = 0)$ describing the sliding motion of the active agent along the wall. When $\tau_r>1$ the linear stability of the fixed point depends on the respective values of all the parameters, but can be summarized as follows. For a given value of the damping coefficients $\nu$ and $\tau_n$, there is a critical inertia above which the fixed point turns unstable in favor of periodic dynamics through a Hopf bifurcation (Fig.~\ref{fig:wall_align}-a,c), leaving the place for the oscillations observed experimentally ((Fig.~\ref{fig:wall_align}-f). Both translational and angular inertia must be nonzero for this instability to take place. Increasing the damping coefficients simply increases the value of the critical inertia. For a fixed ratio of angular to translational inertia, the dependence of the critical inertia on $\tau_r$ is not necessarily monotonic ((Fig.~\ref{fig:wall_align}-e).
\begin{figure}[t!]
\centering
\includegraphics[width=0.99\linewidth]{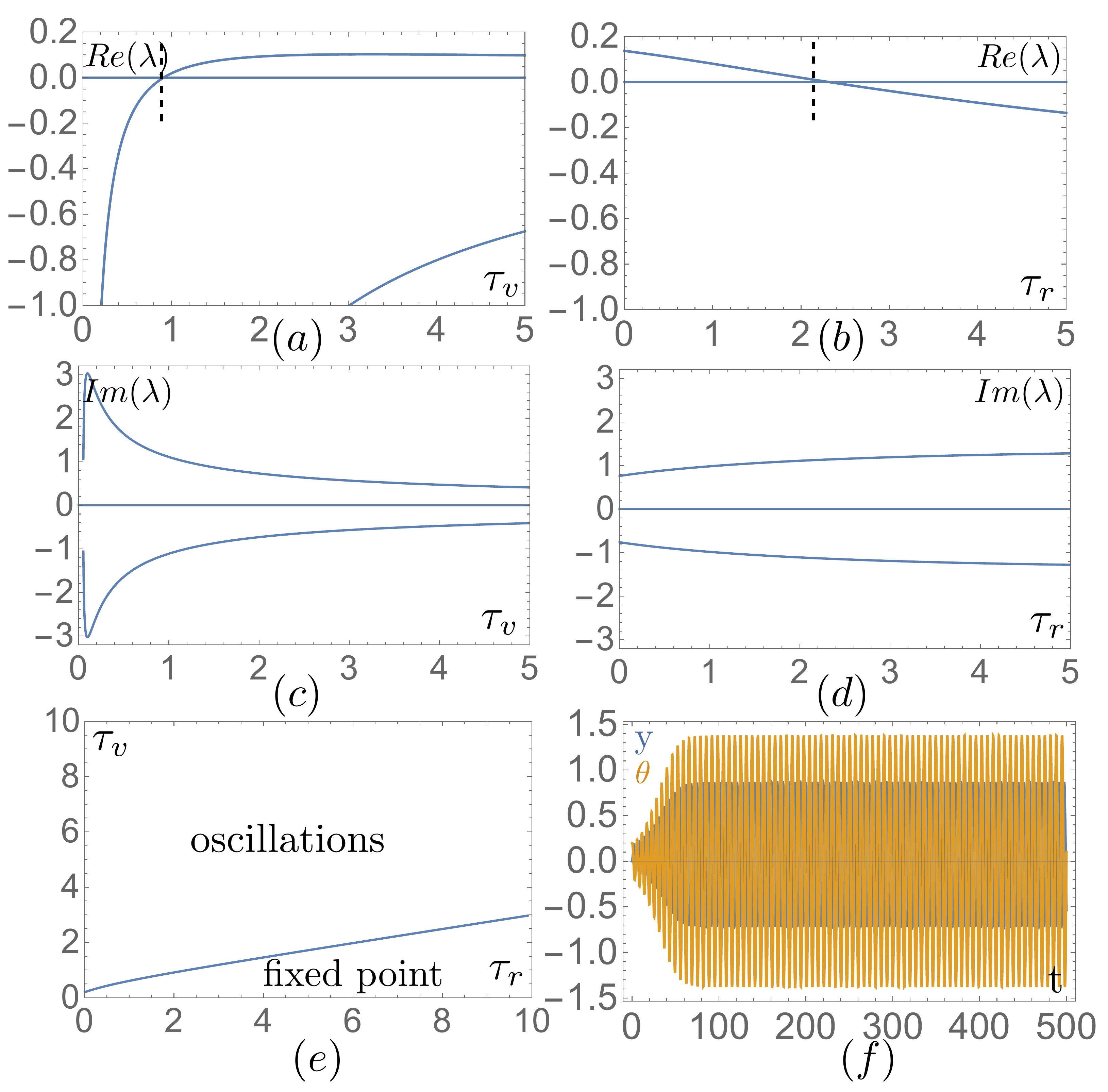}
\caption{{\bf Linear stability of the fixed point against the wall in the anti-aligning case, $\epsilon = -1$:} (a-c) Real and imaginary part of some, including the largest, eigenvalues as a function of $\tau_v$ for $\tau_r=2$: for large enough inertia, the fixed point facing the wall turns unstable via a Hopf bifurcation (vertical dotted line); (b-d) Real and imaginary part of the eigenvalues as a function of $\tau_r$ for $\tau_v=1$ : for large enough friction with the wall, the fixed point facing the wall turns unstable via a Hopf bifurcation (vertical dotted line); (e) The critical value of $\tau_v$ above which oscillations take place is a simple increasing function of $\tau_r$. (f) The oscillating dynamics for $\tau_v=1$ and $\tau_r=2$. In all panels $J = \tau_v,  \tau_n = 0.1, \nu = 0.1$.}
\label{fig:wall_anti}
\end{figure}
In the anti-aligning case, the situation is somehow simpler because the sliding solution does not exist. The 
relevant fixed points remain marginal in the absence of inertia ((Fig.~\ref{fig:wall_anti}-a,c), whatever the value of $\tau_r$. For large enough inertia and not too large $\tau_r$, any of the translationally equivalent fixed point turns unstable via a Hopf bifurcation ((Fig.~\ref{fig:wall_anti}-b,d), leading to the same type of oscillatory dynamics as in the aligning case ((Fig.~\ref{fig:wall_anti}-f). Here the critical inertia is a simple increasing function of $\tau_r$ ((Fig.~\ref{fig:wall_anti}-e). As in the aligning case, increasing the damping coefficients $\nu$ and $\tau_n$ simply increases the value of the critical inertia.

Both in the aligning and anti-aligning cases one numerically checks that the amplitude of the oscillations increases with inertial until eventually $\theta$ reaches values larger than $\pi/2$, the agent leaves the wall and the present simplified description stops holding.

\begin{figure}[t!]
\centering
\includegraphics[width=0.95\linewidth]{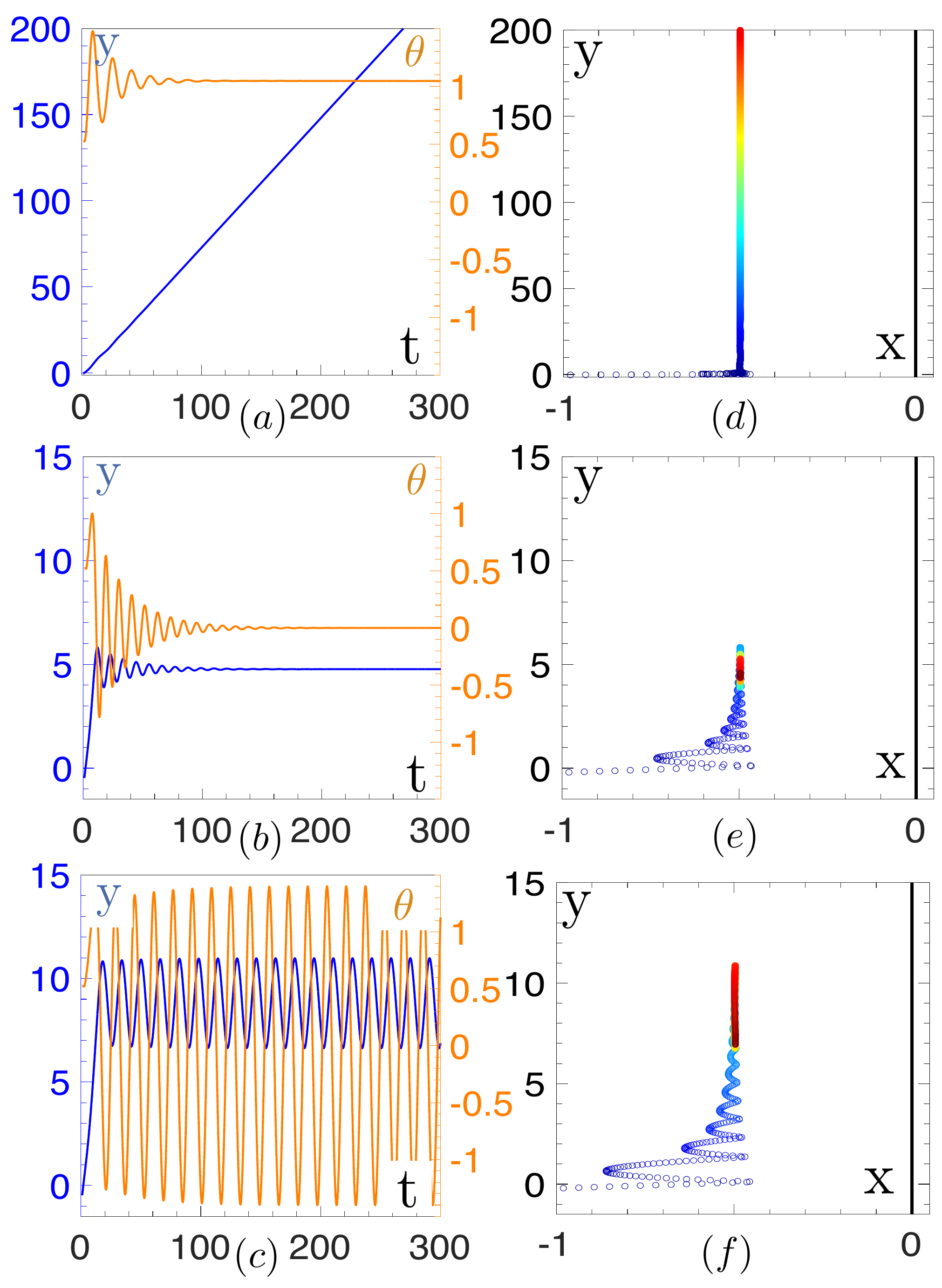}
\caption{{\bf 2d trajectories against the wall in the aligning case, $\epsilon = +1$:} (a-b-c) $y$, coordinate along the wall, and $\theta$, the orientation of the active agent as a function of time for the sliding dynamics, the static fixed point, and the oscillating dynamics respectively; (d-e-f) corresponding trajectories in the $(x,y)$ plane. The vertical black line denotes the position of the wall; the trajectories are color coded from blue to red by increasing time; in the initial condition, the agent sits in $(x=-1, y=-0.5)$ and points in the direction $\theta = \pi/6$. The parameter values are for (a,d): $\tau_v = 0.5, J=4, \tau_r = 0.5, \nu=0.15, \tau_n=0.15$; for (b,e): $\tau_v = 1, J=4, \tau_r = 3, \nu=0.15, \tau_n=0.15$; for (c,f): $\tau_v = 1.5, J=9, \tau_r = 3, \nu=0.15, \tau_n=0.15$.}
\label{fig:2dwall_align}
\end{figure}

\begin{figure}[t!]
\centering
\includegraphics[width=0.95\linewidth]{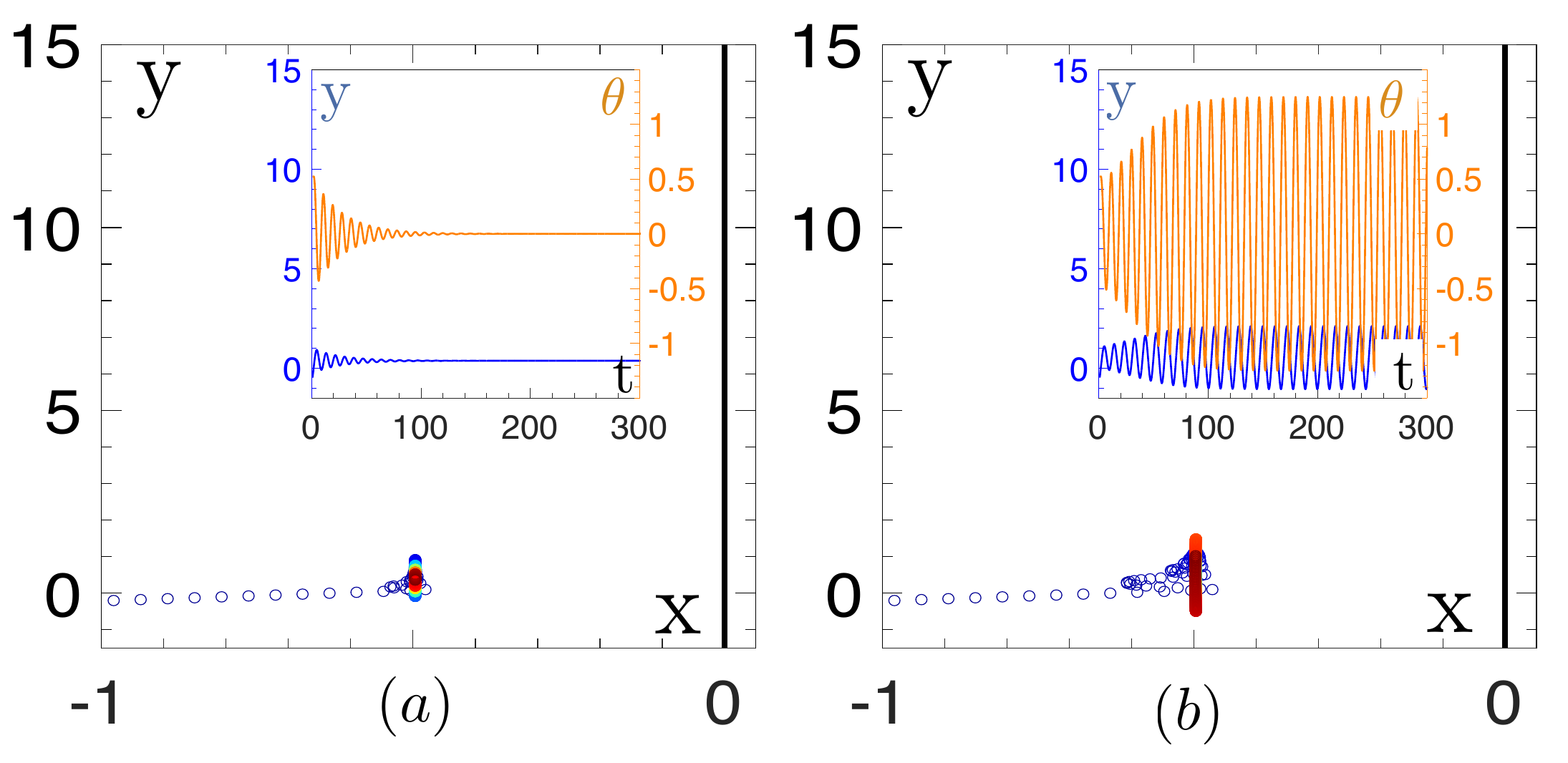}
\caption{{\bf 2d trajectories against the wall in the anti-aligning case, $\epsilon = -1$:} (a-b) $y$, coordinate along the wall, and $\theta$, the orientation of the active agent as a function of time for the static fixed point, and the oscillating dynamics respectively; (insets) corresponding trajectories in the $(x,y)$ plane; the vertical black line denotes the position of the wall; the trajectories are color coded from blue to red by increasing time; in the initial condition the agent sits in $(x=-1, y=-0.5)$ and points in the direction $\theta = \pi/6$. The parameter values are for (a): $\tau_v = 0.2, J=3, \tau_r = 1, \nu=0.15, \tau_n=0.15$; for (b): $\tau_v = 0.5, J=3, \tau_r = 1, \nu=0.15, \tau_n=0.15$.}
\label{fig:2dwall_anti}
\end{figure}

As stated initially the above reduction of the dynamics to a simpler one-dimensional motion along the wall assumes a permanent contact with the wall which is different from the repeated collisions experienced by a real hard body active agent, such as the Vibebot of Fig.~\ref{fig:exp}-(a). Performing simulations of the full set of equations~\ref{eq:Wall_nodim} we confirm the validity of the approach, as the same dynamics are observed in the good range of values for the control parameters. In the aligning case, the three dynamics identified previously are reported in Fig.~\ref{fig:2dwall_align}. For $\tau_r<1$ (Fig.~\ref{fig:2dwall_align}-a,d), the agent indefinitely slides along the wall. This behavior was tested experimentally using an alternative Vibebot with a smaller $\tau_r$ value. Only the sliding dynamics were observed (see Supp. Mat. Movie 4). For $\tau_r>1$ and small inertia (Fig.~\ref{fig:2dwall_align}-b,e), damped oscillations follow an initial bouncing regime, before the agent sets in a static position facing the wall. For $\tau_r>1$ and large inertia (Fig.~\ref{fig:2dwall_align}-c,f), sustained oscillations follow the same initial regime. The two dynamics predicted for the anti-aligning case are also recovered. Independently of the value of $\tau_r$, for small enough inertia (Fig.~\ref{fig:2dwall_anti}-a) the agent rapidly stabilizes into the static fixed point (Fig.~\ref{fig:2dwall_anti}-a), while for larger inertia the oscillations set in (Fig.~\ref{fig:2dwall_anti}-b). Note the difference in the transitory regime, which is much more localized around the impact point in the anti-aligning case, than in the aligning one.

\section{Conclusion}
Coupling inertia with self-alignment considerably enriches the deterministic dynamics of self-propelled active agents.  As in the case of standard active particles, translational inertia hinders the agent's ability to change the direction of its velocity in response to the active and external forces and angular inertia does the same for the direction of the self-propulsion in response to torques. Self-alignment, because of the coupling with the translational degrees of freedom is not simply acting like an active torque. In the aligning case, it reduces the inertial delay between the orientation of self-propulsion and the velocity. Conversely in the anti-aligning case, it increases this delay.

In light of the relevance of inertial self-alignment for large active agents that take their momentum from a substrate, such as walking robots and animals or rolling vehicles, the dynamics discussed here could contribute to better control of such agents. The next step is obviously to consider the role of the noise on such dynamics, following the work of~\cite{Scholz-2018,Lowen-2020}, a technical challenge, given the coupling of positional and translational degrees of freedom.

\bibliography{Active.bib,other.bib}

\end{document}